# Deep Learning Aided Multi-Objective Optimization and Multi-Criteria Decision Making in Thermal Cracking Process for Olefines Production

Seyed Reza Nabavi [a, *], Mohammad Javad Jafari [a], Zhiyuan Wang [b, c]

[a] Department of Applied Chemistry, Faculty of Chemistry, University of Mazandaran, Babolsar, Iran

[b] Department of Chemical and Biomolecular Engineering, National University of Singapore, Singapore 117585, Singapore

[c] AI Research and Computational Optimization (AIRCO) Laboratory, DigiPen Institute of Technology Singapore, Singapore 139660, Singapore

*Corresponding author: srnabavi@umz.ac.ir, Tel: +981135302508, Fax: +981135302340

**Abstract**

**Background:** Multilayer perceptron (MLP) aided multi-objective particle swarm optimization algorithm (MOPSO) is employed in the present article to optimize the liquefied petroleum gas (LPG) thermal cracking process. This new approach significantly accelerated the multi-objective optimization (MOO), which can now be completed within one minute compared to the average of two days required by the conventional approach.

**Methods:** MOO generates a set of equally good Pareto-optimal solutions, which are then ranked using a combination of a weighting method and five multi-criteria decision making (MCDM) methods. The final selection of a single solution for implementation is based on majority voting and the similarity of the recommended solutions from the MCDM methods.

**Significant Findings:** The deep learning (DL) aided MOO and MCDM approach provides valuable insights into the trade-offs between conflicting objectives and a more comprehensive understanding of the relationships between them. Furthermore, this approach also allows for a deeper understanding of the impact of decision variables on the objectives, enabling practitioners to make more informed, data-driven decisions in the thermal cracking process.

**Keywords:** liquefied petroleum gas (LPG); thermal cracking; machine learning (ML); deep learning (DL); multi-criteria decision making (MCDM); multi-objective particle swarm optimization (MOPSO)

1. Introduction

Light olefins, including ethylene and propylene, are critical raw materials in the petrochemical industry, which was valued at USD 556.09 billion in 2021 and is forecasted to experience a compound annual growth rate of 6.2% from 2022 to 2030, according to Grand View Research (2021). These olefins are used in the production of various plastic materials, fibers, and other chemicals (Haribal et al., 2018). They can be produced from light hydrocarbon streams through refining processes like steam cracking (Blay et al., 2017; Karaba et al., 2020; Popelier et al., 2023), which is a significant and energy-intensive process in the industry (Abghari & Sadi, 2013; Amghizar et al., 2017; Feli et al., 2017). During steam cracking, a hydrocarbon feedstock is mixed with steam and cracked at high temperatures in a tubular reactor (Sadrameli, 2015; Van Geem et al., 2008). The feedstock can be a gaseous substance like ethane, propane, or butane, or a liquid like light or heavy naphtha. The process occurs in long, vertical reactors that are heated by gas-fueled burners at high temperatures (between 600 and 850°C) and low pressure. The products of the reaction are cooled in a transfer line exchanger (TLE). Coke formation, which occurs on the inner surface of the cracking coils and in the TLE, can reduce olefin selectivity and increase pressure drop and tube surface temperature. To optimize performance and reduce costs, it is important to predict and optimize the quality of the products through process modeling (Zhou, 2013). However, many real-world optimization problems in chemical engineering involve two or more conflicting objectives and do not have a single optimal solution (Nabavi et al., 2023; Wang et al., 2023). In these cases, multi-objective optimization (MOO) is used to find a set of Pareto-optimal solutions, also known as Pareto-optimal front or non-dominated solutions, which are equally good from the perspective of the given objectives (Datta & Regis, 2016; Rahimi et al., 2018; Hemmat Esfe et al., 2018; Tian et al., 2018). Multi-criteria decision making (MCDM) is then employed to select one solution from the Pareto-optimal front for implementation (Wang et al., 2021; Wang et al., 2022; Park et al., 2023).

There have been several studies on the MOO of thermal cracking plants, including the mathematical modeling of an ethane cracking furnace (Barza et al., 2018), the thermal cracking of an extra-heavy fuel oil (Ghashghaee & Shirvani, 2018), the multi-objective optimization of an ethylene cracking furnace system (Yu et al., 2018; Geng et al., 2016; Yu et al., 2016), and the multi-objective optimization and kinetic modeling of thermal cracking (Taghipour & Naderifar, 2015). It has been observed in the literature that many studies on MOO in thermal cracking process use mathematical models as objective functions. However, these mathematical models are often complex and include a range of mass balance, energy, and momentum equations. As a result, optimization using these models is typically expensive in terms of computational time and cost. Conversely, employing machine learning (ML) or deep learning (DL) models as surrogates to establish complicated relationships between inputs and outputs has recently garnered considerable interest across diverse fields (Wang, Irfan, et al., 2023; Bhoyar et al., 2023; Zhang et al., 2021). The use of ML or DL models can simplify complex processes and reduce the computational costs in optimization (Biegler et al., 2014; Oldani et al., 2015; Choudhury et al., 2017; Irfan et al., 2020; Wang et al., 2022). These surrogate/ML models can then be integrated with MOO, a promising approach for optimizing complex processes (Yu et al., 2021; Ma et al., 2022). For example, Wang et al., (2022) used this approach and tested on two chemical processes: supercritical water gasification process from wet organic wastes to maximize hydrogen production and minimize byproducts (methane, carbon monoxide, and carbon dioxide), and combustion process in a power plant to maximize energy output and minimize emissions of harmful pollutants. The results of these optimizations were favorable.

In the present study, we use multilayer perceptron (MLP) neural networks to model the input-output relationships for the liquefied petroleum gas (LPG) thermal cracking process. The mathematical model of the process is first solved under various operating conditions to gather dataset for training the MLP models. Then, the MLP models are connected to a MOO solver, multi-objective particle swarm optimization (MOPSO) algorithm. This MLP-aided MOPSO is used to optimize the thermal cracking process in six different cases. At the end of each case, the criteria importance through intercriteria correlation (CRITIC) weighting method is used to determine the weight of each objective, and then the following five MCDM methods are applied to select one of the Pareto-optimal solutions for implementation, namely, multi-attributive border approximation area comparison (MABAC), preference ranking on the basis of ideal-average distance (PROBID), simple additive weighting (SAW), sPROBID (simpler PROBID), and technique for order of preference by similarity to ideal solution (TOPSIS).

This comprehensive approach, which systematically integrates DL models, MOO algorithms, weighting techniques, and MCDM methods, remains largely unexplored in the chemical engineering literature. The principal contribution of our current work is to bridge this research gap, thereby accelerating data-driven studies for mathematically complex processes in chemical engineering. As shall be showcased in the later sections, the new approach of DL aided MOO and MCDM significantly reduces the computational time, completing the optimization and analysis within one minute, which is a substantial improvement from the average two-day timeframe required with conventional approach.

The rest of this article is structured as follows. Section 2 presents the methodologies used in the study, including MLP neural network for modeling the input-output relationship, MOPSO algorithm for solving MOO, CRITIC weighting method for finding the weight of each objective, and MABAC, PROBID, SAW, sPROBID, and TOPSIS methods for MCDM. Section 3 demonstrates the application to the thermal cracking process and discusses six case studies. Finally, the conclusions of the present study are summarized in Section 4.

## 2. Methodologies

This section presents the architecture of MLP neural network, the algorithm of MOPSO, the steps of the CRTIC weighting method, and the five MCDM methods, namely, MABAC, PROBID, SAW, sPROBID, and TOPSIS.

### 2.1. Multilayer Perceptron Neural Network

A general architecture MLP artificial neural network consists of three components: the input layer, the hidden layer(s), and the output layer (Li et al., 2017; Irfan & Shafie, 2021). Each layer contains a few neurons, which are interconnected with all the neurons in the following layer. Each neuron in the input layer represents a specific independent variable (i.e., decision variable or input). Similarly, each neuron in the output layer represents a dependent variable (i.e., objective or output) to be predicted (Wang et al., 2022). MLP is a powerful and flexible DL model with a high degree of robustness, capable of mapping nonlinear input-output relationships and handling arbitrarily complex and dynamic systems (Aghbashlo et al., 2015). For example, Fig.S1 in the Supporting Information depicts a simplified architecture of MLP neural network with three neurons in the input layer, one hidden layer with five neurons and one neuron in the output layer. It is worth mentioning that each connection in MLP has a weight associated with it, reflecting its relative importance to the output of the connected neuron in the following layer. The weighted sum of the values of all neurons (plus an optional bias term) in the previous layer that is linked to the current neuron is transformed using an activation function (e.g., sigmoid, relu, or tanh) to determine the output of the current neuron. See Murtagh (1991) for more details about MLP.

## 2.2. Multi-Objective Particle Swarm Optimization

The particle swarm optimization (PSO) algorithm is a powerful heuristic optimization technique used to find the optimal solution to a problem by simulating the social behavior of animals such as a group birds or fish (Niu et al., 2018; Jain et al., 2018; Ullah et al., 2023). The algorithm begins by initializing a population of particles, each of which represents a potential solution to the optimization problem. Each particle has a position vector and a velocity vector, which determine its location and direction of movement within the search space, respectively. As defined in Eqs (1) and (2), the velocity vector of each particle is updated using its current velocity vector, the PBest (personal best position, i.e., the best solution the particle has found so far) and the GBest (global best position, i.e., the best solution found by any particle in the swarm). The position vector of each particle is then updated based on its current position vector and the newly updated velocity vector. This process is repeated iteratively until the optimal solution is found or a predetermined stopping criterion (e.g., number of generations/iterations) is met.

$$V_i^{c+1} = w^c \times V_i^c + a_1^c \times r_1 (PBest_i^c - X_i^c) + a_2^c \times r_2 \times (GBest^c - X_i^c) \qquad (1)$$

$$X_i^{c+1} = X_i^c + V_i^{c+1} \qquad (2)$$

In Eqs (1) and (2) above, $i$ is the index of particle. $c$ is the index of iteration. $X_i^c$ and $V_i^c$ are the position and velocity vectors of the $i^{th}$ particle at the $c^{th}$ iteration, respectively. $r_1$ and $r_2$ are two random numbers distributed uniformly in the range of [0, 1] for velocity update. $w^c$ is the inertia factor at the $c^{th}$ iteration, controlling the influence of previous velocity. $a_1^c$ and $a_2^c$ are the cognitive and social learning factors to balance exploiting the $PBest_i^c$ and $GBest^c$ in the $c^{th}$ iteration, respectively.

The MOPSO algorithm is a variant of the PSO algorithm for solving MOO problems by taking the concept of Pareto dominance into account. MOPSO introduces an external archive set to maintain a fixed number of Pareto-optimal (non-dominated) solutions obtained by the particles. These elite particles are then used as candidates for personal and global best particles. MOPSO also employs a mutation operator to diversify the swarm and improve the global search capability of the particles (Zhang et al., 2017; Niu et al., 2018; Khoshbakht et al., 2023). The seven steps of using MOPSO for MOO are briefly described as follows:

Step 1. Define the computational parameters and randomly generate the initial group of particles in the search space.

Step 2. Calculate the inertia weight and learning coefficients at the current iteration.

Step 3. Evaluate the performance (i.e., optimization objective value) of each particle in the swarm and update the personal best position of each particle, as well as the global best position of the swarm at the current iteration.

Step 4. Update the positions of all particles in the population.

Step 5. Choose non-dominated solutions to update the archive set.

Step 6. Use the mutation operator to enhance the population diversity.

Step 7. Repeat steps 2-6 until the predetermined number of iterations/generations has been reached. When the process is complete, output the Pareto-optimal solutions in the archive set and stop the optimization process.

**2.3. Criteria Importance through Intercriteria Correlation**

The CRITIC weighing method is a technique used to assign weights to the objectives. It is based on the pairwise correlations between objectives in the decision or objective matrix (i.e., the set of Pareto-optimal solutions obtained from MOO) and the standard deviation of each objective. The CRITIC method was introduced by Diakoulaki et al. (1995), who claimed that it produces more balanced weights by considering the information provided by all objectives and facilitating the resolution of trade-offs between the objectives.

The objective matrix contains *m* rows (one row for each solution) and *n* columns (one column for each objective). The objective is either for maximization or minimization. The following symbols are also used in describing the three steps of the CRITIC method; $f_{ij}$: the value of $j^{th}$ objective for the $i^{th}$ solution in the objective matrix; $F_{ij}$: the value of $f_{ij}$ after normalization; $w_j$: the weight assigned to the $j^{th}$ objective.

Step 1. Normalize the objective matrix using the max-min normalization method, as in Diakoulaki et al. (1995).

$$F_{ij} = \frac{f_{ij} - \min_{i \in m} f_{ij}}{\max_{i \in m} f_{ij} - \min_{i \in m} f_{ij}} \text{ for maximization objective} \quad (3)$$

$$F_{ij} = \frac{\max_{i \in m} f_{ij} - f_{ij}}{\max_{i \in m} f_{ij} - \min_{i \in m} f_{ij}} \text{ for minimization objective} \quad (4)$$

Step 2. Calculate the Pearson product moment correlation between any 2 objectives:

$$\rho_{jk} = \frac{\sum_{i=1}^{m}(F_{ij} - \overline{F_j})(F_{ik} - \overline{F_k})}{\sqrt{\sum_{i=1}^{m}(F_{ij} - \overline{F_j})^2}\sqrt{\sum_{i=1}^{m}(F_{ik} - \overline{F_k})^2}} \quad j, k \in [1, n] \quad (5)$$

Here, $\overline{F_j} = \frac{1}{m}\sum_{i=1}^{m} F_{ij}$ and $\overline{F_k} = \frac{1}{m}\sum_{i=1}^{m} F_{ik}$ represent the arithmetic mean of *j*th and *k*th normalized objective, respectively.

Step 3. Compute the standard deviation of each individual normalized objective and then determine the weight for each objective as follows:

$$\sigma_j = \sqrt{\frac{\sum_{i=1}^{m}(F_{ij} - \overline{F_j})^2}{m}} \quad j \in [1, n] \quad (6)$$

$$c_j = \sigma_j \sum_{k=1}^{n}(1 - \rho_{jk}) \quad j \in [1, n] \quad (7)$$

$$w_j = \frac{c_j}{\sum_{k=1}^{n} c_k} \quad j \in [1, n] \quad (8)$$

## 2.4. Multi-Attributive Border Approximation Area Comparison (MABAC)

MCDM is essentially to rank all the Pareto-optimal solutions and select the top-ranked solution. MABAC is a MCDM method proposed by Pamučar & Ćirović (2015). It involves determining the distance between each Pareto-optimal solution and the border approximation area, which is based on the product of the weighted normalized values in each column of objective matrix. To apply the MABAC method, the following steps should be followed: first, construct the weighted normalized objective matrix; second, calculate the border approximation area for each objective; third, compute the distance between each solution and the border approximation areas; fourth, rank the solution based on the distances, with the solution having the largest distance as the top-ranked solution. Due to space constraints, further details of the MABAC method are not included in the present article. For more information on the steps and equations of the MABAC method, refer to Pamučar & Ćirović (2015) or Wang et al. (2020).

## 2.5. Preference Ranking on the Basis of Ideal-average Distance (PROBID)

PROBID is a MCDM method proposed by Wang et al. (2021). The key feature of this method is its comprehensive coverage on the mean solution and different tiers of ideal solutions for ranking purpose. The ideal solutions are comprised of the most positive ideal solution (PIS), 2nd PIS, 3rd PIS, 4th PIS, till the least PIS, that is, the most negative ideal solution (NIS). Distances between each Pareto-optimal solution and these ideal solutions are then computed and combined with the distance from the mean solution to determine the final ranking score of each solution. Due to space constraints, further details of the PROBID method are not included in the present article. See Wang et al. (2021) for the steps and equations of the PRBOID method.

## 2.6. Simple Additive Weighting (SAW)

SAW is a MCDM method introduced by Fishburn (1967) and MacCrimmon (1968). It is one of the simplest MCDM methods. To apply the SAW method, the following steps should be followed; first, construct the weighted normalized objective matrix; second, sum the values of all objectives for each solution; third, select the solution with the highest summation value as the top-ranked solution. Due to space constraints, further details of the SAW method are not included in the present article. See Wang & Rangaiah (2017) for the steps and equations of the SAW method.

**2.7. Simpler Preference Ranking on the Basis of Ideal-average Distance (sPROBID)**

sPROBID (simpler PROBID) is a variant of the PROBID method described in Subsection 2.5. It is simpler than the PROBID method because sPROBID uses the top quarters of the ideal solutions to determine the distance of each Pareto-optimal solution to the positive ideal solutions, and the bottom quarters of the ideal solutions to determine the distance of each Pareto-optimal solution to the negative ideal solutions. The other steps and calculations in sPROBID are similar to those in the PROBID method. See Wang et al. (2021) for the steps and equations of the sPRBOID method.

**2.8. Technique for Order of Preference by Similarity to Ideal Solution (TOPSIS)**

The TOPSIS method is considered the most popular MCDM method has been applied in many fields over the years. It was developed by Hwang & Yoon (1981). In TOPSIS, the selected Pareto-optimal solution is the one having the largest distance to the most NIS and simultaneously having the smallest Euclidean distance to the most PIS. Due to space constraints, further details of the TOPSIS method are not included in the present article. For more information on the steps and equations of the TOPSIS method, refer to Hwang & Yoon (1981) or Wang & Rangaiah (2017).

**3. Application to Thermal Cracking Process**

The thermal cracking process discussed in the present study depend on five process conditions (i.e., inputs or decision variables), which are feed flow rate ($F_{in}$), coil outlet temperature (COT), coil outlet pressure (COP), steam-to-feed ratio (SR), and coil inlet temperature ($T_{in}$). Table 1 shows the range of values that each of these decision variables can take. On the other hand, there are six dependent variables (i.e., outputs or objectives) of the thermal cracking process, namely, annual production of ethylene ($PC_2H_4$), annual production of propylene ($PC_3H_6$), heat duty consumption per year (PQ), reactor run length (RL), ethylene selectivity index ($C_{2=}/C_2$), and severity index ($C_{-3}/C_{3=}$).

Table 1. Value range of the five decision variables for the thermal cracking process

| Decision Variable | $F_{in}$ (ton/h) | COT (°C) | COP (bar) | SR (kgSteam/kgFeed) | $T_{in}$ (°C) |
|---|---|---|---|---|---|
| Range | 10 - 14 | 800 - 853 | 1.5 - 2.5 | 0.2 - 0.4 | 500 - 703 |

Regarding these decision variables, COT plays a crucial role in determining the composition of the cracked gas. Higher COT leads to increased conversion and yield of products such as ethylene. However, higher COT also results in more coke production within the reactor tubes. Increasing SR can decrease hydrocarbon partial pressure, which in turn reduces coke formation; it also increases the temperature and thermal energy of the feed. COP affects selectivity, with higher values resulting in increased reaction compression and coke formation. Higher $T_{in}$ increases the rate of cracking reactions and reduces the thermal energy flux in the reactor (Sadrameli, 2015).

Next, to build MLP neural networks that can model the input-output relationships of the thermal cracking process, data is first generated by running the mathematical model under various operating conditions. As shown in Fig.1, this is done using a complete factorial design with the five input decision variables (i.e., $F_{in}$, COT, COP, SR, and $T_{in}$). Values of the six objective functions/output variables (i.e., $PC_2H_4$, $PC_3H_6$, PQ, RL, $C_{2=}/C_2$, and $C_{-3}/C_{3=}$) are recorded for each set of the operating conditions. The collected dataset is then divided into three sets: training (60%), validation (20%), and test (20%). When training the MLP models, various techniques are implemented such as cross-validation and regularization to ensure that the models generalize well to new, previously unseen data. Moreover, we keep monitoring the performance of the models on the validation set to prevent overfitting during the training phase. After constructing the MLP neural networks for all the objective functions, they are connected to MOPSO (i.e., MLP-aided MOPSO) to perform MOO and obtain a Pareto-optimal front, as shown in the last two steps of Fig.1.

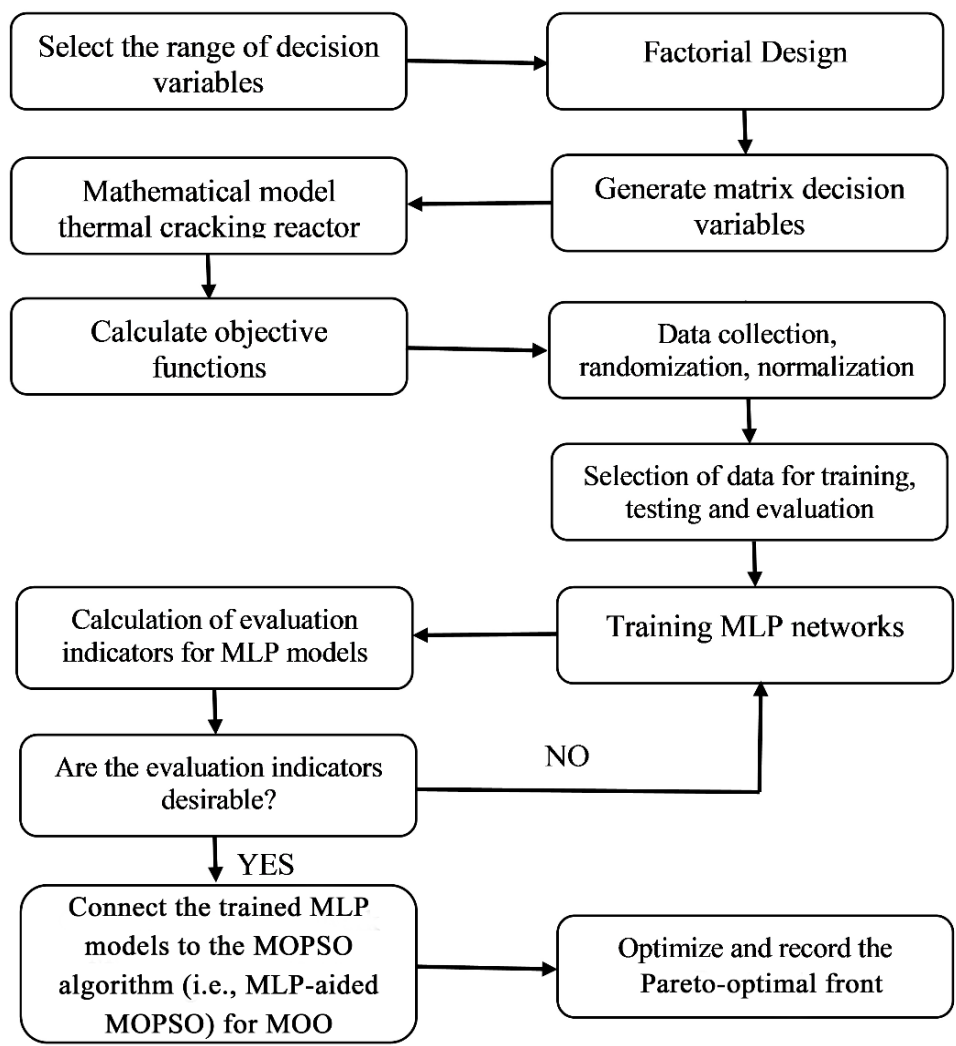

Fig.1. Factorial design for data collection and MLP-aided MOPSO construction

When constructing an MLP model, it is crucial to tune its hyperparameters, such as the number of neurons in the hidden layer(s), the learning rate, and the momentum constant. In this study, we explored a variety of MLP neural network topologies/architectures, utilizing a spectrum of learning rates (ranging from 0.075 to 0.025) and momentum values (ranging from 0.05 to 0.35) to facilitate the prediction of the objective functions. The performance of each model is evaluated using the coefficient of determination ($R^2$) and root mean square error (RMSE) between the actual and predicted values of the objective functions. Fig.2 shows the $R^2$ and RMSE values obtained when experimenting with varying numbers of neurons in the single hidden layer for each objective function, utilizing the test dataset. Given that the number of neurons in the input layer remains constant—equal to the number of input decision variables, which is five—and there is a single neuron in the output layer for each objective, the best MLP architecture for each objective function is then determined by the number of neurons in the single hidden layer that simultaneously yields the highest $R^2$ and the lowest RMSE. It should be noted that our MLP models utilize a single hidden layer. This design choice is grounded in our observation that a single layer can already achieve extremely high $R^2$ values (close to 1) and low RMSE values (close to 0). Introducing additional layers would unnecessarily complicate the models, potentially diminishing their interpretability without offering a commensurate improvement in performance.

Along with the different numbers of neurons in the hidden layer, various values for the learning rate and momentum rate are also evaluated. Table S1 in the Supporting Information presents the optimal MLP architecture and the optimal values for the learning rate and momentum rate for each objective function. The $R^2$ values of the testing dataset for all the objective functions are at least 0.995, showing that all the MLP models perform well. Fig.3 displays scatter plots of the predicted values against the actual values of the test dataset.

In the ideal case, all the points would fall on the diagonal line, meaning that the predicted values equal the actual values. These scatter plots demonstrate that the points are all very close to the diagonal line for each objective function, indicating a high degree of accuracy in the predictions.

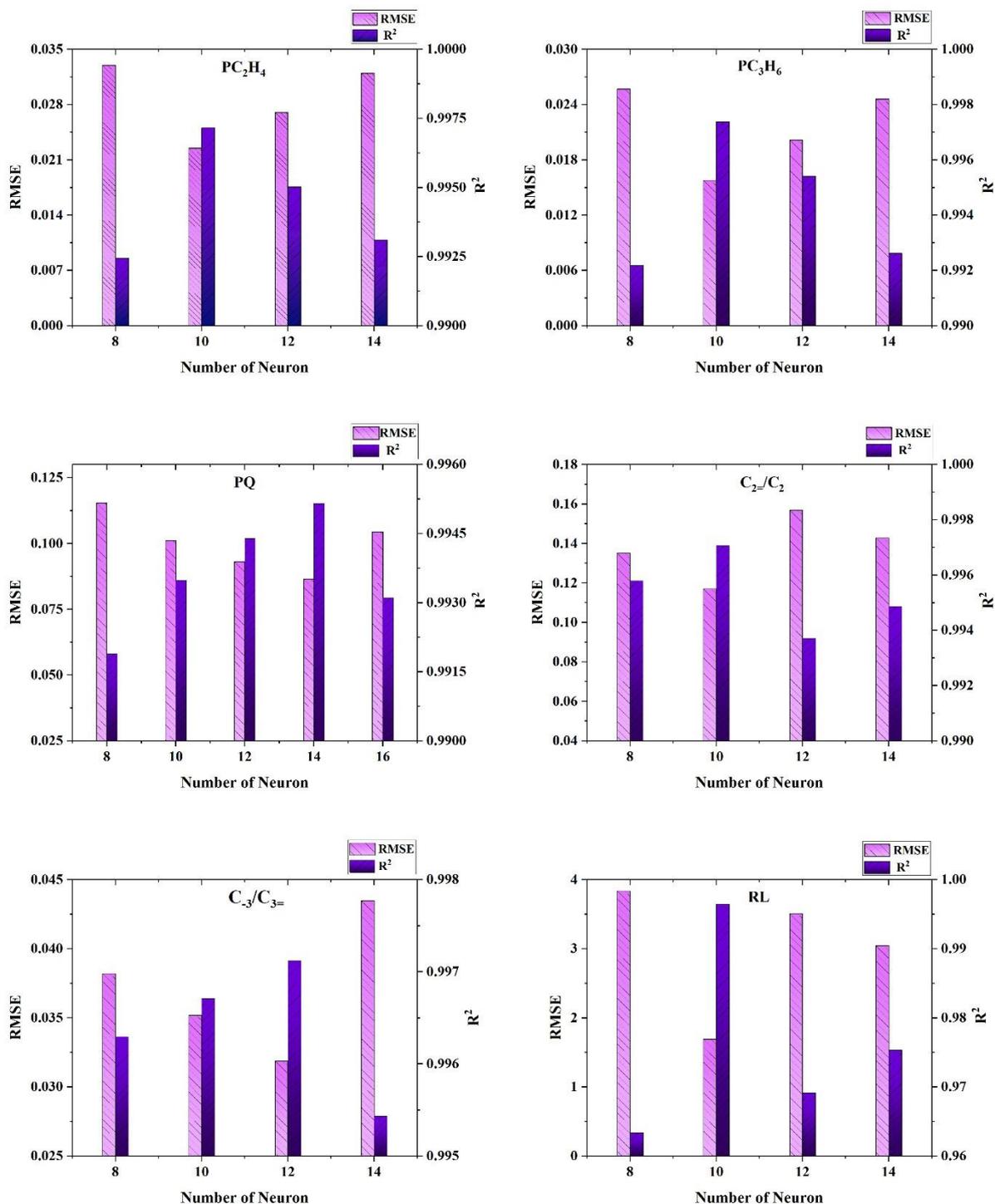

Fig.2. T The $R^2$ and RMSE of different MLP topologies for each objective function when

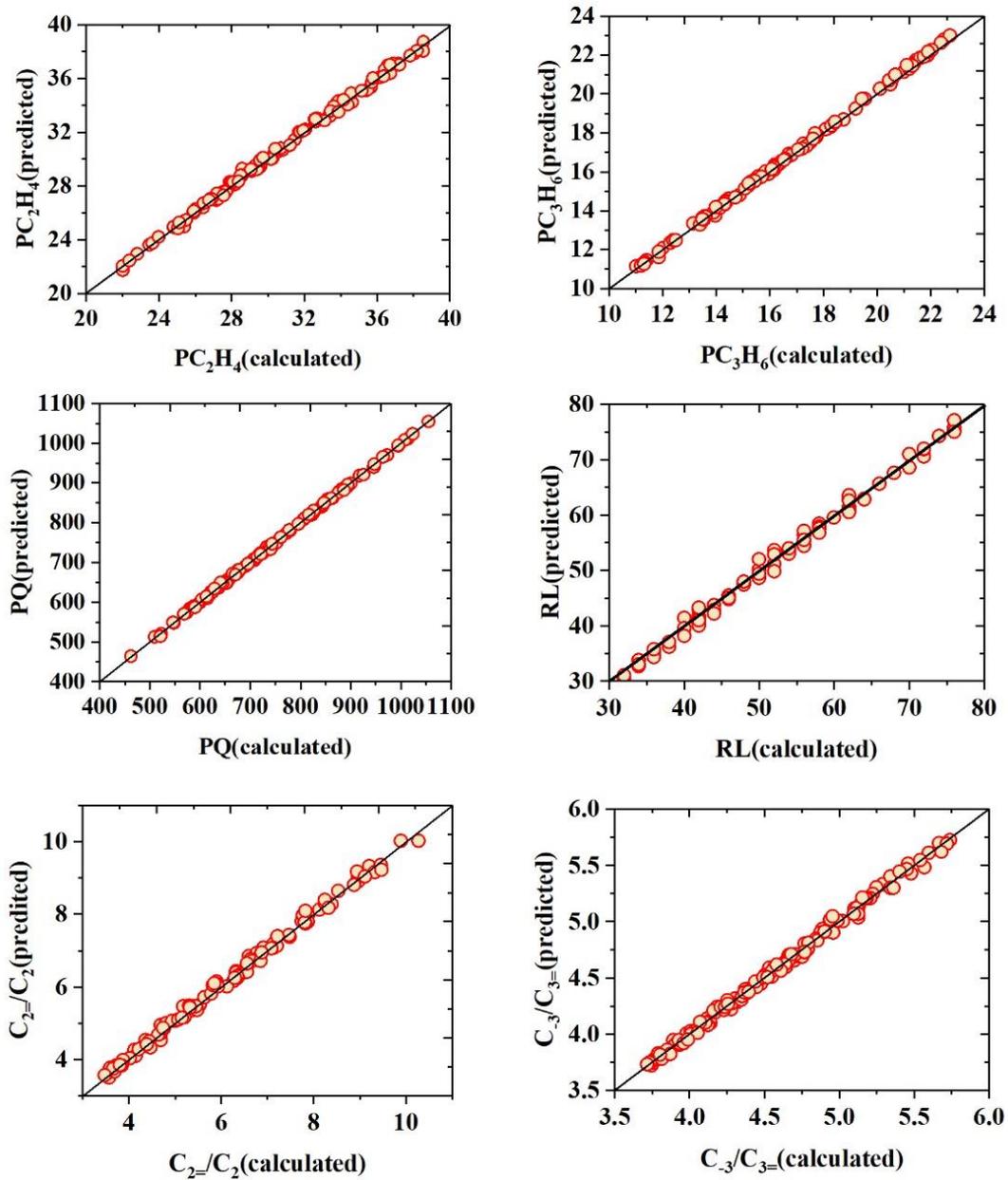

Fig.3. Performance of MLP models on testing dataset: predicted values against actual values

In the following subsections (3.1 to 3.6), six case studies of the thermal cracking process for MOO are investigated and presented.

### 3.1. Case A: Maximizing annual ethylene and propylene production

According to the theory of heat fraction kinetics, an increase in ethylene production leads to a decrease in propylene production (Towfighi et al., 2006). In the present study, simultaneous optimization of these two objectives is investigated. Fig.4 illustrates the Pareto-optimal front for the annual production of ethylene (i.e., $PC_2H_4$) and propylene (i.e., $PC_3H_6$), obtained using MOPSO. As can be seen, the Pareto-optimal front is wide and uniform; besides, as predicted by theory, increasing ethylene production results in a decrease in propylene and vice versa.

It is worth mentioning that the MLP-aided MOPSO was able to complete its optimization within one minute for case A and subsequent cases, using a population of either 60 or 70 particles and running for 100 generations. This is a significant improvement compared to the average time required for optimization utilizing the mathematical model with MOPSO, which takes an average of two days on the same computer.

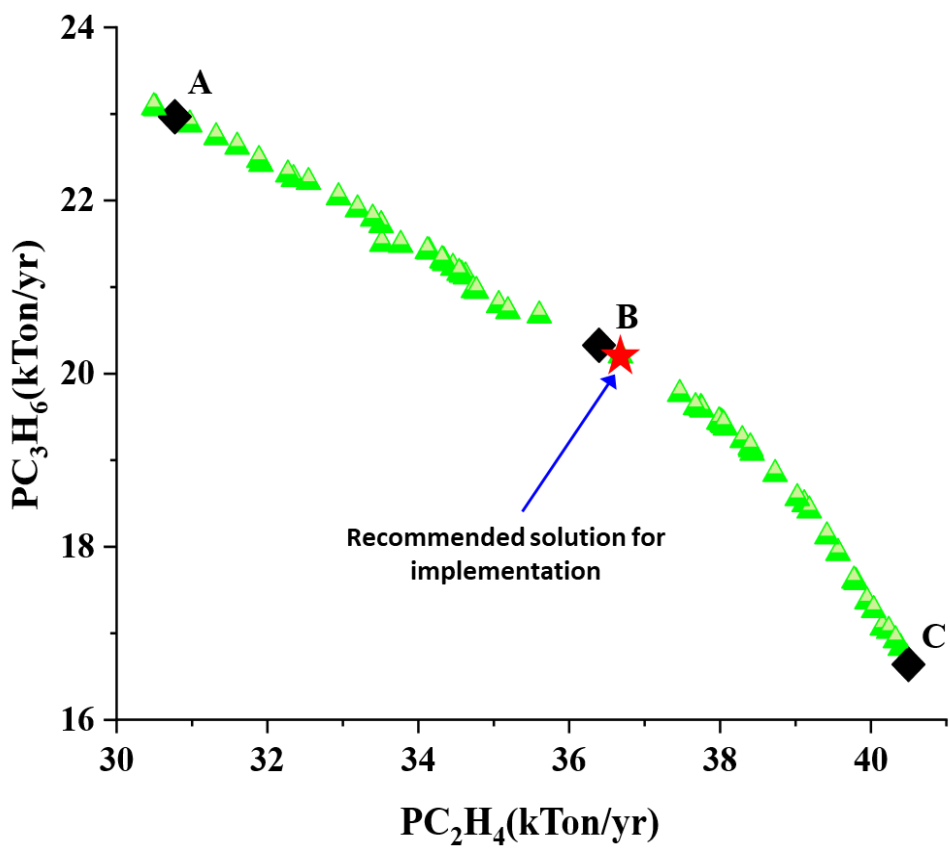

Fig.4. Pareto-optimal front for maximizing the annual production of ethylene and propylene in case A; the recommend solution (★) for implementation after MCDM analysis in case A

Table S2 in the Supporting Information presents in detail the three Pareto-optimal solutions (i.e., points A, B, and C) from Fig.4, along with their corresponding decision variable and objective values.

Fig.S2 in the Supporting Information shows the behavior of decision variables (i.e., $F_{in}$, COT, COP, SR, $T_{in}$) in relation to the production of ethylene (i.e., $PC_2H_4$). As depicted in Fig.S2, $F_{in}$ is chosen at its maximum value for all times to boost the production of both products. $T_{in}$ is also chosen at its highest value as increasing the inlet temperature increases the speed of heat fractions, which is desirable for achieving higher production of both products. Additionally, COT and SR have a positive relationship with $PC_2H_4$, meaning that as one increases, the other also tends to increase. This is because higher COT can increase the rate of the fraction reaction and improve ethylene production; higher SR can decrease the partial feed pressure, which is also beneficial for increasing ethylene production. From another perspective, higher COT can intensify the formation of coke reactions. To mitigate this issue, SR is raised from 0.2 to 0.4, as shown in Fig.S2(d). On the other hand, from the angle of propylene production, low COT, high COP, and low SR are more desirable as they can increase propylene production, which contrasts with ethylene production.

To further ensure the robustness of the results obtained using MLP, we selected four solutions (including the values of their corresponding decision variables) produced by the MLP-aided MOPSO and then compared them to results (i.e., ethylene and propylene production) obtained through directly solving mathematical model. Fig.S2(b) shows the four solutions (i.e., I, II, III, and IV on the plot), which are represented by triangles with different orientations. The comparison results presented in Table 2 demonstrate that the MLP model can produce outputs that are highly similar to those obtained using the mathematical model. This further verification process allowed us to confidently assert the reliability and validity of the results provided by the MLP-aided MOPSO before we applied to more complex case studies subsequently.

Table 2. Comparison results between the MLP model and mathematical model in case A

|  |  | I | II | III | IV |
|---|---|---|---|---|---|
| Decision Variables | $F_{in}$ (ton/h) | 14.00 | 13.98 | 14.00 | 14.00 |
|  | COT (°C) | 817.96 | 816.70 | 812.70 | 811.54 |
|  | COP (bar) | 2.08 | 2.28 | 1.95 | 2.29 |
|  | SR (kgsteam/kgfeed) | 0.20 | 0.38 | 0.20 | 0.40 |
|  | $T_{in}$ (°C) | 703 | 703 | 703 | 703 |
| MLP Model Outputs | $PC_2H_4$ (kton/yr) | 34.4 | 34.5 | 33.2 | 33.3 |
|  | $PC_3H_6$ (kton/yr) | 21.2 | 21.1 | 21.7 | 21.6 |
| Mathematical Model Outputs | $PC_2H_4$ (kton/yr) | 34.1 | 34.1 | 33.0 | 33.0 |
|  | $PC_3H_6$ (kton/yr) | 20.8 | 20.8 | 21.4 | 21.4 |

Remember that Pareto-optimal solutions are all considered equally good because it is impossible to improve any one objective without negatively impacting at least one other objective (e.g., in case A, increasing $PC_2H_4$ inevitably causes a decrease of $PC_3H_6$ as Fig.4 depicts). Nevertheless, only one solution needs to be selected from the Pareto-optimal front for implementation. Therefore, in the present study, five different MCDM methods, namely, MABAC, PROBID, SAW, sPROBID, and TOPSIS, are applied to rank the Pareto-optimal solutions and recommend one of them. The weights for the two objectives, $PC_2H_4$ and $PC_3H_6$, are determined using the CRITIC weighting method and are 0.519 and 0.481, respectively. Table 3 presents the recommended solutions (including values of decision variables and objectives) by the five MCDM methods. As observed from the gray shaded columns in Table 3, PROBID, sPROBID, and TOPSIS recommend the same solution from the Pareto-optimal front. MABAC and SAW recommend two different solutions; however, these solutions are not significantly different from the one selected by PROBID, sPROBID, and TOPSIS, with comparable $PC_2H_4$ and $PC_3H_6$. Overall, based on majority voting and similarity with those chosen by MABAC and SAW, the solution selected by PROBID, sPROBID, and TOPSIS is finally recommended for implementation. Fig.4 shows the recommended solution as a red star (★) on the Pareto-optimal front of case A.

Table 3. Recommended Pareto-optimal solutions by five MCDM methods in case A

|  |  | MABAC | PROBID | SAW | sPROBID | TOPSIS |
|---|---|---|---|---|---|---|
| Decision Variables | $F_{in}$ (ton/h) | 14 | 14 | 14 | 14 | 14 |
|  | COT (ºC) | 831.7 | 824.5 | 830.2 | 824.5 | 824.5 |
|  | COP (bar) | 1.5 | 1.5 | 1.5 | 1.5 | 1.5 |
|  | SR (kgsteam/kgfeed) | 0.393 | 0.387 | 0.392 | 0.387 | 0.387 |
|  | $T_{in}$ (ºC) | 703 | 703 | 703 | 703 | 703 |
| Objectives | $PC_2H_4$ (kton/yr) | 38.01 | 36.68 | 37.75 | 36.68 | 36.68 |
|  | $PC_3H_6$ (kton/yr) | 19.44 | 20.21 | 19.60 | 20.21 | 20.21 |

## 3.2. Case B: Maximizing annual ethylene production and minimizing heat duty consumption per year

Heat duty consumption (i.e., PQ) is a key consideration in the thermal cracking process because of its impact on profitability. Fig.5(a) presents the Pareto-optimal front for maximizing annual ethylene production and minimizing heat duty consumption per year. To facilitate a comparison of cases A and B, PQ of case A was calculated and compared to that of case B in Fig.5(b); it reveals that the range of PQ against $PC_2H_4$ for case B is larger than for case A, and that case B has lower heat duty consumption than case A at the same level of ethylene production.

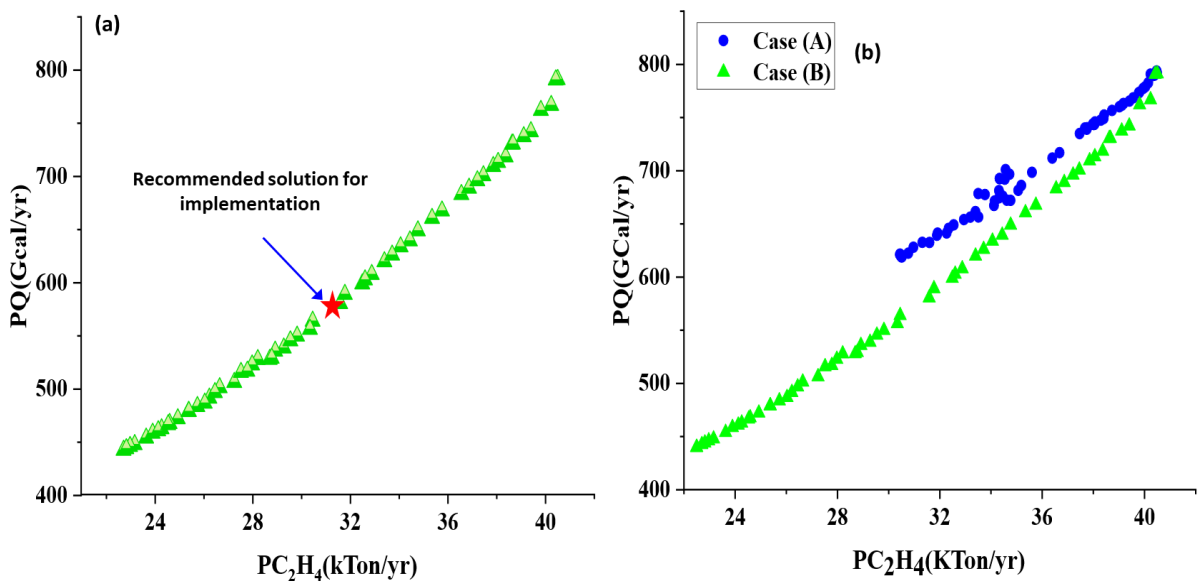

Fig.5. Pareto-optimal front for maximizing ethylene production and minimizing heat duty consumption in case B - (a); the recommend solution (★) for implementation after MCDM analysis in case B; comparison between cases A and B - (b)

Fig.S3 in the Supporting Information shows the behavior of decision variables (i.e., $F_{in}$, COT, COP, SR, $T_{in}$) in relation to the production of ethylene (i.e., $PC_2H_4$) for case B. As depicted in Fig.S3(b), the point at which COT has its lowest value (800°C) corresponds to the lowest $PC_2H_4$. From here, COT rapidly increases through MOPSO to boost ethylene production until the $PC_2H_4$ reaches approximately 28 kton/yr (as indicated by the vertical dashed line in Fig.S3(b). At the same time, $F_{in}$ remains relatively stable before the vertical dashed line in Fig.S3(a), maintaining within a range of 10 to 11 ton/h. On the other hand, after $PC_2H_4$ hitting the 28 kton/yr mark, COT starts to level off, aiming to slow down the increase in heat duty consumption, while $F_{in}$ starts to rise rapidly to continue increasing the $PC_2H_4$. In addition, it is known from theory that increasing the SR increases the feed rate inside the reactor, resulting in an increased heat duty consumption to reach the reaction temperature. As a result, SR tends to fluctuate around a small value of 0.25 in case B. Furthermore, $T_{in}$ is chosen at the maximum value, which can reduce the energy consumption of the reactor.

Like in case A, only one solution is required to be selected from the Pareto-optimal front for implementation. The five MCDM methods, namely, MABAC, PROBID, SAW, sPROBID, and TOPSIS, are applied to rank the Pareto-optimal solutions of case B and recommend one of them. The weights for the two objectives, $PC_2H_4$ and PQ, are determined using the CRITIC weighting method and are 0.511 and 0.489, respectively. Table 4 presents the recommended solutions (including values of decision variables and objectives) by the five MCDM methods. As observed from the gray shaded columns in Table 4, sPROBID and TOPSIS recommend the same solution from the Pareto-optimal front. MABAC, PROBID, and SAW recommend different solutions. Overall, based on majority voting and the similarity of the recommended solutions, the one selected by sPROBID and TOPSIS is finally recommended for implementation. Fig.5(a) shows the recommended solution as a red star (★) on the Pareto-optimal front of case B.

Table 4. Recommended Pareto-optimal solutions by five MCDM methods in case B

|  |  | MABAC | PROBID | SAW | sPROBID | TOPSIS |
|---|---|---|---|---|---|---|
| Decision | $F_{in}$ (ton/h) | 13.20 | 11.65 | 13.77 | 11.45 | 11.45 |

| Variables | COT (°C) | 853.0 | 846.9 | 852.7 | 846.8 | 846.8 |
|---|---|---|---|---|---|---|
| | COP (bar) | 1.58 | 1.66 | 1.58 | 1.70 | 1.70 |
| | SR (kgsteam/kgfeed) | 0.260 | 0.252 | 0.248 | 0.246 | 0.246 |
| | $T_{in}$ (°C) | 703 | 703 | 703 | 703 | 703 |
| Objectives | $PC_2H_4$ (kton/yr) | 38.37 | 32.48 | 39.11 | 31.59 | 31.59 |
| | PQ (gcal/yr) | 718.45 | 598.95 | 737.37 | 580.44 | 580.44 |

## 3.3. Case C: Maximizing annual ethylene and propylene production and minimizing heat duty consumption per year

In case C, we examine the impact of heat duty consumption on the annual ethylene and propylene production by maximizing $PC_2H_4$ and $PC_3H_6$ while minimizing PQ concurrently. Fig.6 illustrates the Pareto-optimal front of case C in a 3D plot.

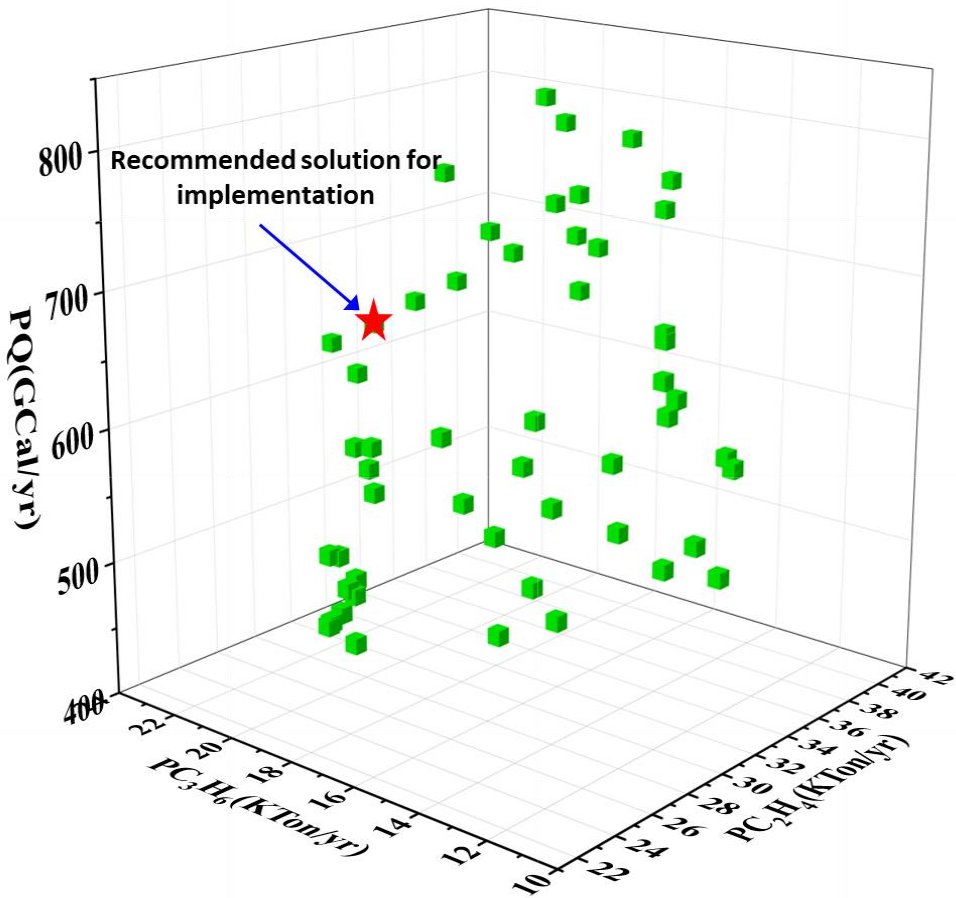

Fig.6. Pareto-optimal front for maximizing annual ethylene and propylene production and minimizing heat duty consumption per year in case C; the recommend solution (★) for The results obtained are then compared to those of cases A and B. In case A, the two-objective optimization was focused on maximizing $PC_2H_4$ and $PC_3H_6$. In case B, the two-objective optimization aimed to maximize $PC_2H_4$ and minimize PQ. As shown in Fig.7, when producing the same amount of ethylene, case A requires the highest heat duty in general, as it did not consider heat duty in the optimization. In contrast, case B typically requires the lowest heat duty, as minimizing heat duty was one of the two objectives for optimization. Regarding case C, the heat duty values are more varied, with results generally falling between those of cases A and B, as heat duty is one of the three objectives considered in the optimization. Based on the results of these three cases, it can be concluded that the three-objective optimization in case C is effective and produces acceptable results.

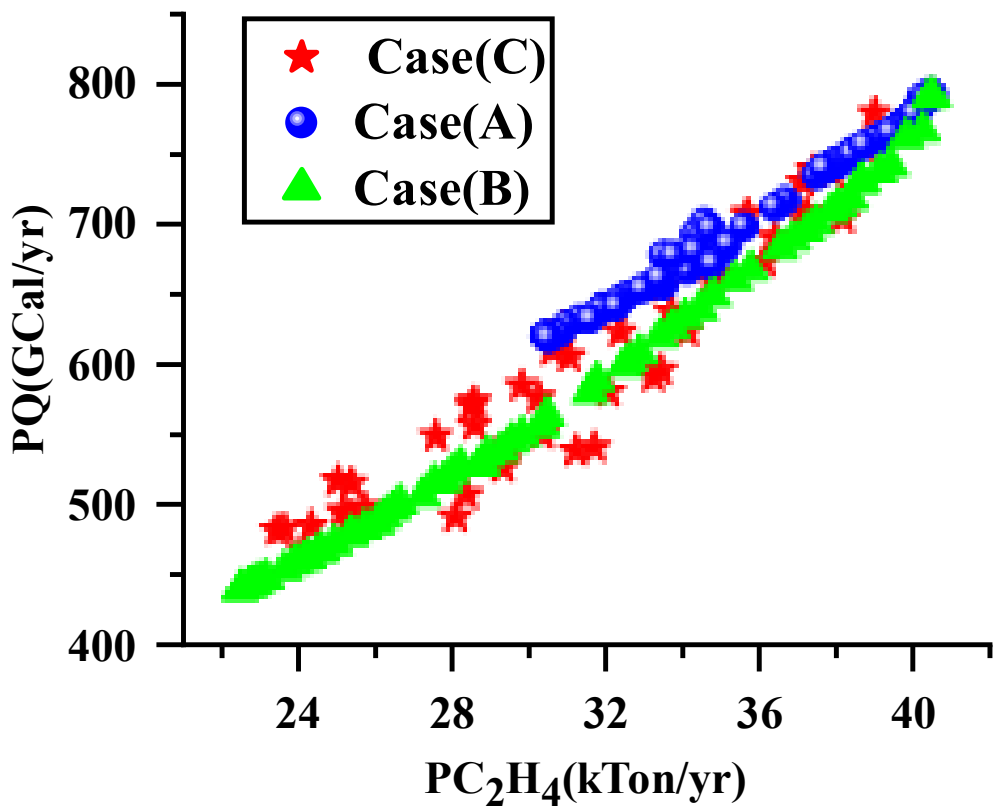

Fig.7. Comparison of cases A, B, and C for heat duty (PQ) against ethylene production ($PC_2H_4$)

Fig.S4 in the Supporting Information shows the behavior of decision variables (i.e., $F_{in}$, COT, COP, SR, $T_{in}$) in relation to the production of ethylene (i.e., $PC_2H_4$) for case C. As previously discussed in cases A and B, higher $F_{in}$, higher COT, lower COP, higher SR, and higher $T_{in}$ are beneficial for increasing ethylene production. As before, one solution needs to be selected from the Pareto-optimal front for implementation. The five MCDM methods, namely, MABAC, PROBID, SAW, sPROBID, and TOPSIS, are applied to rank the Pareto-optimal solutions of case C and recommend one of them. The weights for the three objectives, $PC_2H_4$, $PC_3H_6$, and PQ, are determined using the CRITIC weighting method and are 0.383, 0.291, and 0.326, respectively. Table 5 presents the recommended solutions (including values of decision variables and objectives) by the five MCDM methods. As observed from the gray shaded columns in Table 5, SAW, sPROBID, and TOPSIS recommend the same solution from the Pareto-optimal front. MABAC and PROBID recommend different solutions. Overall, based on majority voting and the similarity of the recommended solutions, the one selected by SAW, sPROBID, and TOPSIS is finally recommended for implementation. Fig.6 shows the recommended solution as a red star (★) on the Pareto-optimal front of case C.

Table 5. Recommended Pareto-optimal solutions by five MCDM methods in case C

| | | MABAC | PROBID | SAW | sPROBID | TOPSIS |
|---|---|---|---|---|---|---|
| Decision Variables | $F_{in}$ (ton/h) | 14.00 | 14.00 | 13.97 | 13.97 | 13.97 |
| | COT (ºC) | 827.8 | 811.5 | 807.0 | 807.0 | 807.0 |
| | COP (bar) | 1.72 | 2.50 | 2.46 | 2.46 | 2.46 |
| | SR (kgsteam/kgfeed) | 0.237 | 0.378 | 0.254 | 0.254 | 0.254 |
| | $T_{in}$ (ºC) | 703 | 703 | 703 | 703 | 703 |
| Objectives | $PC_2H_4$ (kton/yr) | 36.40 | 33.50 | 32.24 | 32.24 | 32.24 |
| | $PC_3H_6$ (kton/yr) | 19.94 | 21.60 | 22.27 | 22.27 | 22.27 |

| | PQ (gcal/yr) | 689.93 | 655.30 | 639.65 | 639.65 | 639.65 |

## 3.4. Case D: Maximizing annual ethylene and propylene production and the run length

The run length (RL) is another important factor in the thermal fraction process, as the formation of coke on the reactor walls can eventually lead to reduced performance and require the furnace to be taken removed from the production line for cleaning. As a result, increasing the run length can be beneficial for increasing production. In other words, a longer run length allows the heat fraction process to continue without interruption, leading to higher overall production. Therefore, the three-objective optimization for maximizing $PC_2H_4$, $PC_3H_6$, and RL is investigated in case D and its results are compared to case A. Fig.8 shows the Pareto-optimal front of case D in a 3D plot. As observed, the RL value can vary from 30 to 76 days.

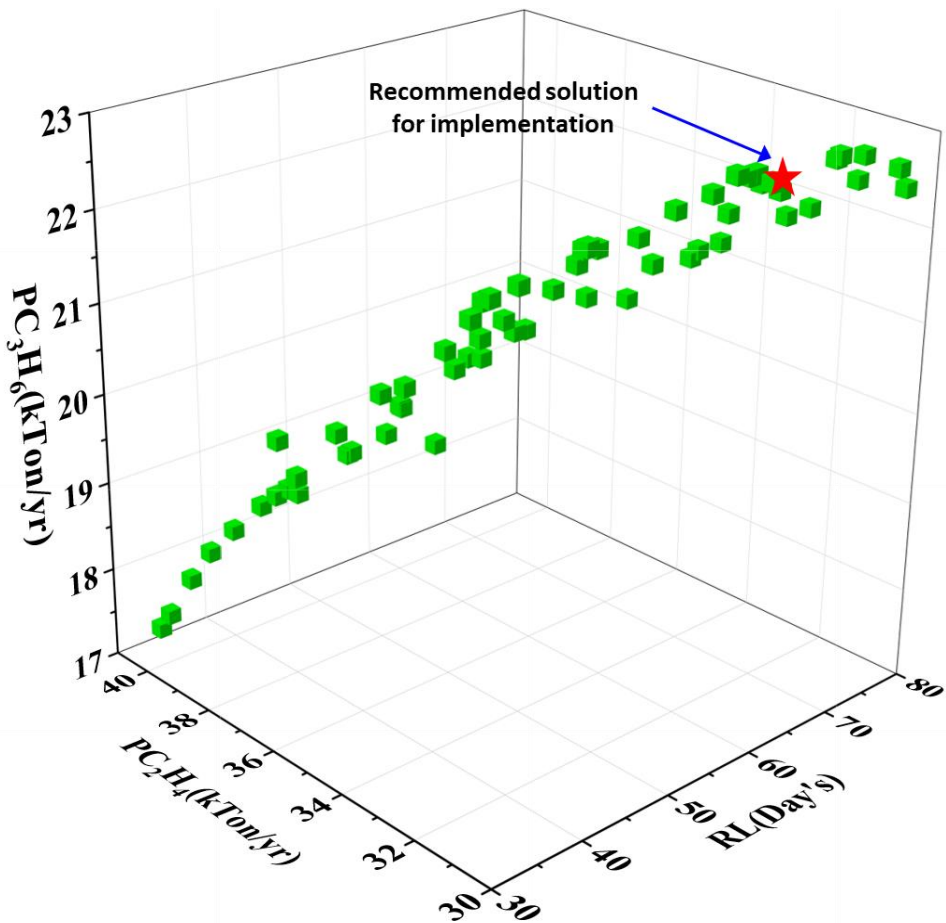

Fig.8. Pareto-optimal front for maximizing annual ethylene and propylene production and the run length in case D; the recommend solution (★) for implementation after MCDM analysis in case D

Fig.S5 in the Supporting Information compares the results of cases A and D. As shown in Fig.S5(a), the plots for $PC_3H_6$ against $PC_2H_4$ are largely overlapping between the two cases, with case D generally having slightly lower $PC_3H_6$ at the same level of $PC_2H_4$, particularly when $PC_2H_4$ is less than 38 kton/yr. At the same time, Fig.S5(b) shows that the RL of case D is generally higher than or equivalent to that of case A. In other words, it can be said that at the same level of $PC_2H_4$, an increase in RL comes at the expense of a decrease in $PC_3H_6$. The three-objective optimization in case D provides valuable insights into the trade-off between the three objectives, allowing for a more nuanced and in-depth understanding of the relationships between them.

As before, it is not practical to implement all the Pareto-optimal solutions, so a single solution needs to be selected from the Pareto-optimal front. The five MCDM methods, namely, MABAC, PROBID, SAW, sPROBID, and TOPSIS, are applied to rank the Pareto-optimal solutions of case D and recommend one of them. The weights for the three objectives, $PC_2H_4$, $PC_3H_6$, and RL, are determined using the CRITIC weighting method and are 0.471, 0.265, and 0.264, respectively. Table 6 presents the recommended solutions (including values of decision variables and objectives) by the five MCDM methods. As observed from the gray shaded columns in Table 6, PROBID, sPROBID, and TOPSIS recommend the same solution from the Pareto-optimal front. MABAC and SAW recommend different solutions. Overall, based on majority voting and the similarity of the recommended solutions, the one selected by PROBID, sPROBID, and TOPSIS is finally recommended for implementation. Fig.8 shows the recommended solution as a red star (★) on the Pareto-optimal front of case D.

Table 6. Recommended Pareto-optimal solutions by five MCDM methods in case D

|  |  | MABAC | PROBID | SAW | sPROBID | TOPSIS |
|---|---|---|---|---|---|---|
| Decision Variables | $F_{in}$ (ton/h) | 14.00 | 13.97 | 13.95 | 13.97 | 13.97 |
|  | COT (°C) | 806.8 | 810.4 | 808.9 | 810.4 | 810.4 |
|  | COP (bar) | 1.50 | 1.50 | 1.50 | 1.50 | 1.50 |
|  | SR (kgsteam/kgfeed) | 0.200 | 0.297 | 0.273 | 0.297 | 0.297 |
|  | $T_{in}$ (°C) | 703 | 703 | 643 | 703 | 703 |
| Objectives | $PC_2H_4$ (kton/yr) | 31.75 | 32.57 | 31.69 | 32.57 | 32.57 |
|  | $PC_3H_6$ (kton/yr) | 22.15 | 21.48 | 21.65 | 21.48 | 21.48 |
|  | RL (days) | 73.42 | 71.86 | 75.96 | 71.86 | 71.86 |

**3.5. Case E: Maximizing the ethylene selectivity index and minimizing the severity index**

The ethylene selectivity index (i.e., $C_{2=}/C_2$) is defined as the ratio of the ethylene fraction in the outlet to the ethane fraction in the inlet. The severity index (i.e., $C_{-3}/C_{3=}$) is defined as the sum of the fractions of propane, propylene, propadiene, ethane, ethylene, acetylene, methane, and hydrogen, divided by the fraction of propylene. Fig. 9 shows the Pareto-optimal front of case E for maximizing the ethylene selectivity index and minimizing the severity index simultaneously.

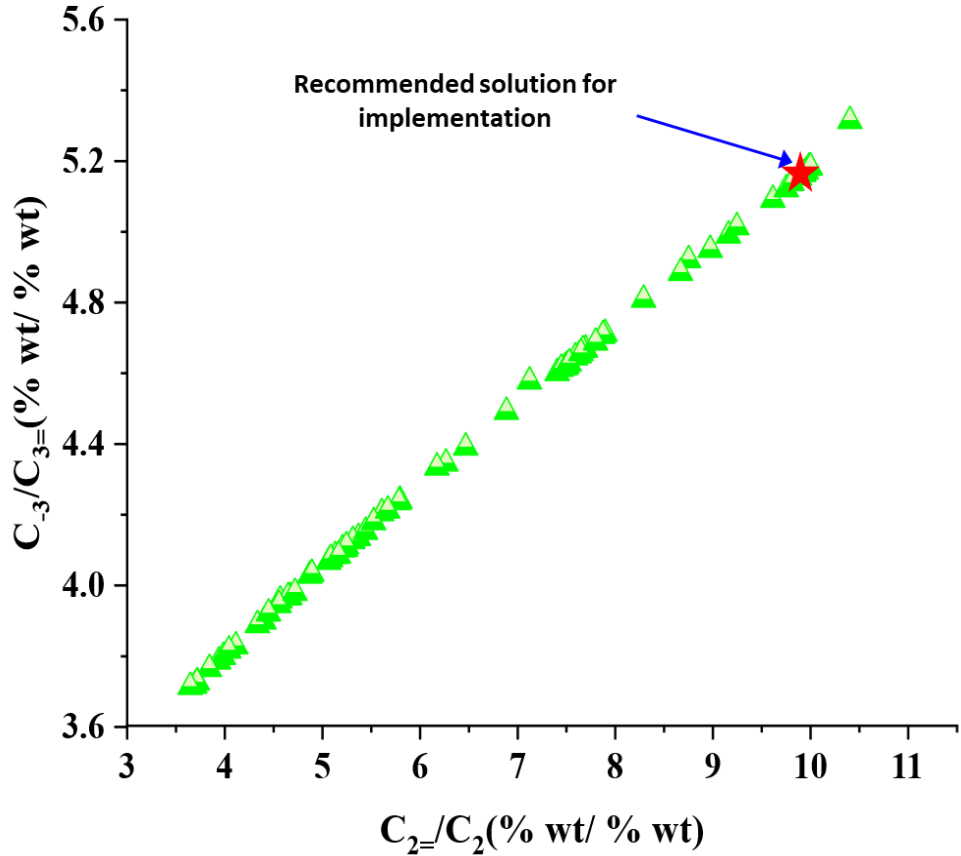

Fig. 9. Pareto-optimal front for maximizing the ethylene selectivity index and minimizing the severity index in case E; the recommend solution (★) for implementation after MCDM analysis in case E

Fig.S6 shows the behavior of decision variables (i.e., $F_{in}$, COT, COP, SR, $T_{in}$) in relation to the ethylene selectivity index (i.e., $C_{2=}/C_2$) for case E. As depicted in Fig.S6(b), the ethylene selectivity index increases as COT increases and becomes almost constant in the end. On the other hand, Fig.S6(a) shows that $F_{in}$ is relatively stable at the beginning, fluctuating between 13 and 14 ton/h, before plummeting to approximately 10 ton/h toward the end. Additionally, the remaining three subplots of Fig.S6 demonstrate that lower COP, higher SR, and lower $T_{in}$ are all advantageous for achieving a higher ethylene selectivity index.

As before, one solution needs to be selected from the Pareto-optimal front for implementation. The five MCDM methods, namely, MABAC, PROBID, SAW, sPROBID, and TOPSIS, are applied to rank the Pareto-optimal solutions of case E and recommend one of them. The weights for the two objectives, $C_{2=}/C_2$ and $C_{-3}/C_{3=}$, are determined using the CRITIC weighting method and are 0.507 and 0.493, respectively. Table 7 presents the recommended solutions (including values of decision variables and objectives) by the five MCDM methods. As observed from the gray shaded columns in Table 7, MABAC and PROBID recommend the same solution from the Pareto-optimal front. SAW, sPROBOD, and TOPISIS recommend different solutions. Overall, based on majority voting and the similarity of the recommended solutions, the one selected by MABAC and PROBID is finally recommended for implementation. Fig. 9 shows the recommended solution as a red star (★) on the Pareto-optimal front of case E.

Table 7. Recommended Pareto-optimal solutions by five MCDM methods in case E

| | | MABAC | PROBID | SAW | sPROBID | TOPSIS |
|---|---|---|---|---|---|---|
| Decision Variables | $F_{in}$ (ton/h) | 13.35 | 13.35 | 11.62 | 13.51 | 13.11 |
| | COT (ºC) | 853 | 853 | 853 | 853 | 853 |
| | COP (bar) | 1.50 | 1.50 | 1.50 | 1.50 | 1.50 |
| | SR (kgsteam/kgfeed) | 0.40 | 0.40 | 0.40 | 0.40 | 0.40 |
| | $T_{in}$ (ºC) | 500.0 | 500.0 | 500.0 | 500.6 | 500.0 |
| Objectives | $C_{2=}/C_2$ (%wt / %wt) | 9.93 | 9.93 | 10.40 | 9.89 | 10.00 |
| | $C_{-3}/C_{3=}$ (%wt / %wt) | 5.16 | 5.16 | 5.31 | 5.15 | 5.18 |

**3.6. Case F: Maximizing the ethylene selectivity index and the run length and minimizing the severity index**

In case F, the maximization of ethylene selectivity index (i.e., $C_{2=}/C_2$) and the run length (i.e., RL) and the minimization of the severity index (i.e., $C_{-3}/C_{3=}$) are investigated, and the results are compared with those of case E. Fig.10 presents the Pareto-optimal front of case F.

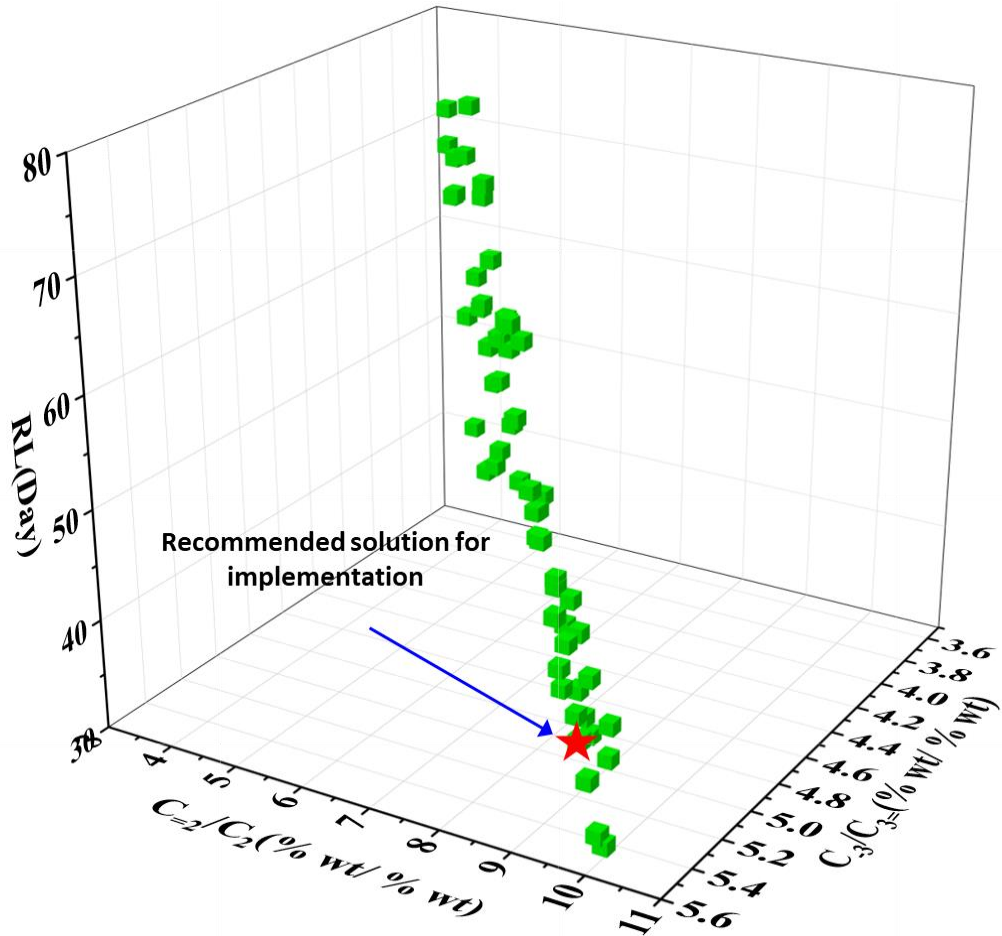

Fig.10. Pareto-optimal front for maximizing the ethylene selectivity index and the run length and minimizing the severity index in case F; the recommend solution (★) for implementation after MCDM analysis in case F

To facilitate a comparison of cases E and F, the run length for case E was calculated and compared to that of case F in Fig.S7 of the Supporting Information. As can be seen from Fig.S7(a), the plots for $C_{-3}/C_{3=}$ against $C_{2=}/C_2$ are partially overlapping between the two cases, with case F generally having slightly higher $C_{-3}/C_{3=}$ at the same level of $C_{2=}/C_2$, particularly when $C_{2=}/C_2$ is greater than 6. On the other hand, Fig.S7(b) shows that the RL of case F is generally higher than or equivalent to that of case E. In other words, it can be said that at the same level of $C_{2=}/C_2$, an increase in RL comes at the expense of an increase in $C_{3-}/C_{3=}$. The three-objective optimization in case F provides valuable insights into the trade-off between the three objectives, allowing for a more nuanced and in-depth understanding of the relationships between them.

Fig.S8 shows the behavior of decision variables (i.e., $F_{in}$, COT, COP, SR, $T_{in}$) in relation to the ethylene selectivity index (i.e., $C_{2=}/C_2$) for case F. Like the analysis in case E, lower $F_{in}$, higher COT, lower COP, higher SR, and lower $T_{in}$ are all beneficial for a higher ethylene selectivity index.

As before, one solution needs to be selected from the Pareto-optimal front for implementation. The five MCDM methods, namely, MABAC, PROBID, SAW, sPROBID, and TOPSIS, are applied to rank the Pareto-optimal solutions of case F and recommend one of them. The weights for the three objectives, $C_{2=}/C_2$, RL, and $C_{3-}/C_{3=}$, are determined using the CRITIC weighting method and are 0.473, 0.268, and 0.259, respectively. Table 8 presents the recommended solutions (including values of decision variables and objectives) by the five MCDM methods. As observed from the gray shaded columns in Table 8, PROBID, sPROBID, and TOPSIS recommend the same solution from the Pareto-optimal front. MABAC and SAW recommend different solutions. Overall, based on majority voting and the similarity of the recommended solutions, the one selected by PROBID, sPROBID, and TOPSIS is finally recommended for implementation. Fig.10 shows the recommended solution as a red star (★) on the Pareto-optimal front of case F.

Table 8. Recommended Pareto-optimal solutions by five MCDM methods in case F

|  |  | MABAC | PROBID | SAW | sPROBID | TOPSIS |
|---|---|---|---|---|---|---|
| Decision Variables | $F_{in}$ (ton/h) | 13.69 | 10.00 | 10.00 | 10.00 | 10.00 |
|  | COT (ºC) | 812.5 | 842.9 | 847.8 | 842.9 | 842.9 |
|  | COP (bar) | 1.53 | 1.50 | 1.50 | 1.50 | 1.50 |
|  | SR (kgsteam/kgfeed) | 0.36 | 0.40 | 0.40 | 0.40 | 0.40 |
|  | $T_{in}$ (ºC) | 503.3 | 500.0 | 503.9 | 500.0 | 500.0 |
| Objectives | $C_{2=}/C_2$ (%wt / %wt) | 4.82 | 9.36 | 10.04 | 9.36 | 9.36 |
|  | RL (days) | 75.80 | 37.52 | 31.53 | 37.52 | 37.52 |
|  | $C_{-3}/C_{3=}$ (%wt / %wt) | 4.05 | 5.35 | 5.50 | 5.35 | 5.35 |

## 4. Conclusions

In this study, MLP-aided MOPSO algorithm was used to optimize the LPG thermal cracking process in six case studies. The mathematical model of the thermal cracking process was first solved under various operating conditions to generate a dataset for training MLP models. These trained MLP models performed well on the testing dataset and had a high degree of accuracy in their predictions, which were then connected to MOPSO for MOO. The MLP-aided MOPSO greatly reduced the computational time required for MOO in all six case studies, completing the optimization within one minute, which is a considerable improvement compared to the average of two days needed when using the mathematical model with MOPSO on the same computer. The MOO process resulted in a Pareto-optimal front, representing a set of equally good solutions. To select a single solution from the Pareto-optimal front for implementation, the CRITIC weighting method was used to assign weights to the objectives; then, five MCDM methods, MABAC, PROBID, SAW, sPROBID, and TOPSIS, were applied to rank all the Pareto-optimal solutions, each recommending one solution. The final selection for implementation was based on majority voting and the similarity of the recommended solutions. Overall, these DL aided MOO and MCDM analyses not only provide valuable insights into the trade-off between conflicting objectives, but they also allow for a more comprehensive and detailed understanding of the relationships between these objectives. In addition, these analyses enable us to fully grasp the impact of decision variables on the objectives and make more informed data-driven decisions in the thermal cracking process.

## Nomenclature

*Abbreviations*

| | |
|---|---|
| COP | coil outlet pressure |
| COT | coil outlet temperature |
| CRITIC | criteria importance through intercorrelation |
| DL | deep learning |
| LPG | liquefied petroleum gas |
| MABAC | multi-attributive border approximation area comparison |
| MCDM | multi-criteria decision making methods |
| ML | machine learning |
| MLP | multilayer perceptron |
| MOO | multi-objective optimization |
| MOPSO | multi-objective particle swarm optimization algorithm |
| PROBID | preference ranking on the basis of ideal-average distance |
| RMSE | root mean square error |
| SAW | simple additive weighting |
| SR | steam-to-feed ratio |
| TOPSIS | technique for order of preference by similarity to ideal solution |
| TLE | transfer line exchanger |